\documentclass{ws-ijmpa}
\begin{document}

\markboth{}
{Polarisation Transfer in Hyperon Photoproduction at MAMI}

%
\catchline{}{}{}{}{}
%

\title{MEASUREMENT OF POLARISATION TRANSFER IN HYPERON PHOTOPRODUCTION AT MAMI}

\author{T.C JUDE, D.I. GLAZIER, D.P. WATTS for the CrystalBall@MAMI and A2 collaborations.}

\address{School of Physics, University of Edinburgh, King's Buildings\\
Edinburgh, EH9 3JZ,
United Kingdom
t.jude@ed.ac.uk}

\maketitle

\begin{history}
\received{Day Month Year}
\revised{Day Month Year}
\end{history}

\begin{abstract}
The photoproduction of $K^{+}$ mesons is an important challenge to recent QCD based chiral perturbation theories in the strange quark sector and is an important constraint on the nucleon excitation spectrum. We present preliminary data from a new high precision measurement using the Crystal Ball detector. The measurement pioneers a new technique for tagging strangeness using detailed cluster analysis in segmented calorimeters which has potential wider application at present and future hadron physics facilities.

\end{abstract}

\ccode{PACS numbers: 11.25.Hf, 123.1K}

\section{Hyperon Photoproduction and Polarisation Observables}

The photoproduction of a pseudo scalar meson such as the kaon can be described by four complex amplitudes which result in sixteen real observables \cite{Barker}.  Measurement of at least eight well chosen observables gives a "complete measurement" which is sufficient to fully constrain these photoproduction amplitudes. Currently this is an important goal of the leading photon beam facilities and measurements with combinations of polarised beam, polarised targets and $\Lambda$ recoil polarisation are being carried out, with the main physics motivation to better establish the nucleon excitation spectrum. As well as resonance information, accurate data in the threshold region is needed to challenge recent theoretical models based on effective field theories. 

There is presently no detailed measurement of the $\gamma p \rightarrow K^{+} \Lambda$ in the threshold region. At higher energies new quality data is available from Jlab \cite{BradfordCrossSec} and ELSA\cite{Glander}. However, significant discrepancies exist between the cross sections from the two previous measurements for photon energy ranges accessible at MAMI \cite{Mart}. The JLab measurement also extracted the total polarisation transferred to the $\Lambda$ after photoproduction with a circular polarised photon beam. This observable was close to one for all measured kinematics and for photon beam energies 1.032 to 2.741~GeV which has led to speculation about the reaction mechanism \cite{Bradford}.

We present preliminary spectra from the $\gamma p \rightarrow K^{+} \Lambda$ reaction with a circular polarised tagged photon beam in the energy range 0.9-1.4~GeV using the Crystal Ball at MAMI-C.

\section{The Crystal Ball Detector at the Mainz Microtron}

The Mainz Microtron (MAMI-C) is an electron accelerator facility in Mainz, Germany, capable of accelerating electrons up to 1.5~GeV.  The electrons produce circularly or linearly polarised photons via bremsstrahlung in a solid radiator. The photon energy is tagged by momentum analysis of the recoiling bremsstrahlung electrons in the Glasgow tagger \cite{Cameron}.

The Crystal Ball \cite{Starostin} is a calorimeter with 672 optically isolated NaI crystals.  The output of the photomultiplier for each crystal is digitised using multi-hit TDC and ADC modules.  Surrounding the liquid hydrogen target at the centre of the Crystal Ball is the Particle Identification Detector (PID).  The PID comprises 24 plastic scintillator detectors parallel to the beam and read out by photomultipliers at the upstream end.  $\Delta E-E$ analysis provides a means of proton, charged pion and electron identification.  The Crystal Ball is an excellent neutral meson detector via the detection of the decay photons.

TAPS \cite{Novotny} is a 350 element BaF$_{2}$ spectrometer which covers the forward angle range.  The Crystal Ball and TAPS cover approximately 93\% of 4$\pi$ steradian. The combined setup is shown in Fig. \ref{fig:CBPic}. 

\section{Identification of Strange Decay Channels}

It is not possible to identify $K^{+}$ using the PID detector alone as the particles have mass and charge too similar to the proton to be effectively separated using $\Delta E-E$ techniques and the subsequent weak decay of the particles in the detectors spoils the unique energy signal. In previous measurements large magnetic fields and tracking detectors were employed to circumvent this problem. For the case of the Crystal Ball this was impractical due to the geometrical constraints of the detector. A new technique was therefore developed by our group to realise these important strangeness measurements.

\begin{figure}
\begin{center}
\fbox{\includegraphics[angle=0.,width=7cm]{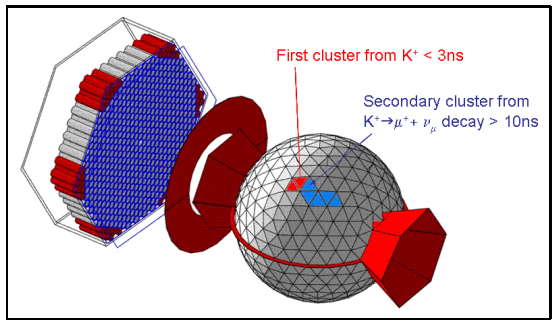}}
\caption{Schematic diagram of identifying $K^{+}$ via their decay in the Crystal Ball.  The red shaded crystals detect the stopped $K^{+}$.  The blue shaded crystals detect the decay of the $K^{+}$ between a timing gate of 10-50 ns after the stopped $K^{+}$.\label{fig:CBPic}}
\end{center}
\end{figure}

\begin{figure}
\begin{center}
\fbox{\includegraphics[angle=0.,width=12cm]{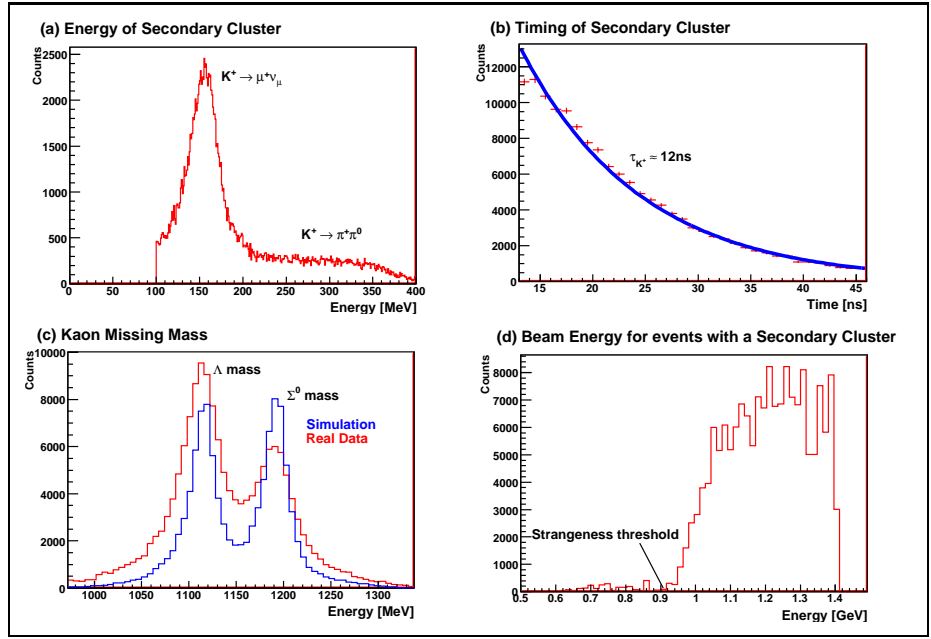}}
\caption{(a)~Energy and (b)~timing of the secondary clusters. (c)~Missing mass and (d)~the photon beam energy from Identified $K^{+}$.\label{fig:StrangeMedley}}
\end{center}
\end{figure}

The basic principle of the technique is illustrated in Fig. \ref{fig:CBPic}. The $K^{+}$ meson enters the Crystal Ball where it is stopped. The crystals involved in this initial interaction are identified as those having a timing within 3~ns of the reaction time. The meson subsequently can decay weakly via:  $K^{+} \rightarrow \mu^{+} \nu_{\mu}$ or  $K^{+} \rightarrow \pi^{+} \pi^{0}$ processes with branching ratios 64\% and 36\% respectively and a half life of 12~ns.  The characteristic energy deposition arising from these processes can clearly be identified in the energy spectra (Fig. \ref{fig:StrangeMedley}(a)). The timing of the delayed weak decay can be resolved using the 2~ns timing resolution of the Crystal Ball detector elements. In fact the half life of the kaon can be extracted from observing the count rate as a function of time. The result compares favourably with the accepted value (Fig. \ref{fig:StrangeMedley}(b)). The crystals in the secondary cluster are identified with a timing gate of between 10-50~ns later than the kaon entry time.  A GEANT4 simulation is used to optimise upper and lower limits on parameters such as the number of crystals in the cluster from the stopped $K^{+}$ and the $K^{+}$ decay, the distance to the furthest crystal and the total energy deposition.  A subsequent $\Delta E - E$ analysis using the PID provides a final constraint to suppress background from charged particles.

Fig. \ref{fig:StrangeMedley}(c) shows the reconstructed missing mass of the system recoiling from the $K^{+}$ which has clear peaks at the masses of the $\Lambda$ and $\Sigma^{0}$, and Fig. \ref{fig:StrangeMedley}(d) is the energy of the photon beam for events in which a $K^{+}$ decay cluster occurred.  The number of events below the threshold for strangeness production at 911~MeV averages very close to zero. The Geant4 simulations of the detection process indicate the efficiency for the $K^{+}$ decay cluster method is approximately 8\%.

\section{Asymmetry Measurements}

\begin{figure}
\begin{center}
\fbox{\includegraphics[angle=0.,width=9.5cm]{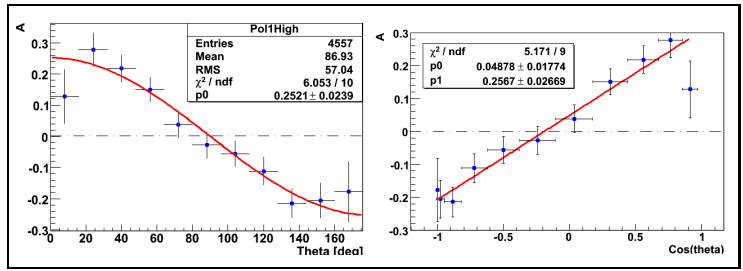}}
\caption{Asymmetry and the cosine of the asymmetry for the polar angle of the decay $\Lambda \rightarrow n \pi^{0}$ and $E_{\gamma}$ below 1.2~GeV. \label{fig:Asymm}}
\end{center}
\end{figure}

The $\Lambda$ decays to either $\Lambda \rightarrow p \pi^{-}$ or $\Lambda \rightarrow n \pi^{0}$ with branching ratios approximately 64~\% and 36~\% respectively.  These are weak decays and parity violating.  As a result, the $\Lambda$ polarisation can be measured via the distribution of its decay particles.  Fig. \ref{fig:Asymm} is the constructed beam helicity asymmetry, $A$, as a function of the polar angle of the decay neutron from the $\Lambda$. The plot shows clearly the expected \nolinebreak{cos$\theta$} modulation. These spectra will be analysed to extract the polarisation transfer observables Cx and Cz.

\section{Conclusion}
We have developed a new techique for tagging strangeness which enables new accurate measurements of strangeness photoproduction using the Crystal Ball at MAMI-C. Further production data will be obtained in the future to improve the statistical precision of the data set. The MAMI measurement will provide accurate data in the threshold region and the new measurements at higher Egamma will help to resolve present discrepancies between the world data.

\end{document}